\documentclass[a4paper,11pt]{article}
\usepackage[utf8]{inputenc}
\usepackage[T1]{fontenc}
\usepackage{epsfig}
\usepackage{float}
\begin{document} 

\begin{center}
  {\LARGE \bf Longitudinal profiles of Extensive Air Showers with inclusion of charm and bottom particles}
\end{center}

\vspace{0.9cm}

\begin{center}
{\large M. A. Müller$^{1,2}$ and V. P. Gonçalves$^1$}\\
{\small $^1$ Instituto de Física e Matemática, Departamento de Física e Matemática, 
Universidade Federal de Pelotas/UFPEL, Pelotas, RS, Brazil}\\
{\small $^2$ Instituto de Física Gleb Wataghin (IFGW), Departamento de Cronologia, Raios Cósmicos, 
Altas Energias e Léptons (DRCC), Universidade Estadual de Campinas, Campinas, SP - Brazil}
\end{center}

\vspace{1.1cm}

\begin{center}
{\large \bf Abstract}
\end{center}

{\small Charm and bottom particles are rare in Extensive Air Showers but the effect of its presence can be radical in the development of the Extensive Air Showers (EAS). 
If such particles arise  with a large fraction of the primary energy, they can reach large atmospheric depths, 
depositing its energy in deeper layers of the atmosphere. As a consequence, the EAS observables ($X_{max}$, $RMS$ and $N_{max}$) will be modified, as well as the shape of the longitudinal profile of the energy deposited in the 
atmosphere. In this paper, we will modify the CORSIKA Monte Carlo by the inclusion of charm and bottom production  in the first interaction of the primary cosmic ray. Results for different selections of the typical  $x_F$ values of the heavy particles  
and distinct production models will be presented.}

\vspace{1cm}

\section{Introduction}

The description of the Extensive Air Showers (EASs) is fundamental for the Cosmic Ray Physics. Primary particles reache the Earth with energies 
up to $10^{20}\;eV$. Such energies are well above those reached in colliders, and, therefore, the simulations of these 
EASs require an extrapolation of the known physics. 

Very energetic charm and bottom heavy hadrons may be produced in the upper atmosphere when a primary cosmic ray or a
leading hadron in an EASs collide with the air. Because of its short mean life, $\approx 10^{-12}\;s$ ($\approx 300\; \mu m$), 
heavy hadrons decay before interacting. At $E\approx 10^{7}\;GeV$ heavy hadrons reach their critical energy 
and its decay probabilities decrease rapidly. Decay lenghts grow to considerable values. At $E\approx 10^{8}\;GeV$ 
their decay lenght becomes of order $\approx10\;km$, implying that they tend to interact in the air instead of decaying. 
Since  the inelasticity in these collisions is much smaller than the one in proton and pion collisions, keeping a higher 
fraction of their energy after each interaction, there could be rare events where a heavy hadron particle transports a 
significant amount of energy deep into the atmosphere, giving rise to additional contributions to the EAS development. 
Heavy particles can be produced at any stage of the EAS development, but it is mainly during the first interaction of the 
primary collision that they are produced with a significant fraction of primary energy.

The collisions of heavy hadrons with the air are very elastic. For example, a $D$ meson after a $10^{9}\;GeV$ collision could keep 
around 55\% of the initial energy, whereas a $B$ meson will have typically 80\% of the incident energy after colliding with 
an air nucleus. In contrast, the leading meson after a $10^{9}\;GeV$ pion collision would carry in average just 22\% of the 
energy \cite{garcia_canal_art_aires}. If heavy hadrons are produced with a high fraction of the primary particle, they will 
interact rather than decaying. If several elastic interactions occur, these heavy hadrons could transport a 
significat amount of energy deep into the atmosphere and likely have observable effects on the EAS development. 

The energy deposition of a heavy partycle near of the ground would produce muons and other particles that could 
change significantly the EASs longitudinal profile seen in fluorescence telescopes (e.g. in the Pierre Auger Observatory \cite{Auger}) and/or the temporal distribution observed in the surface detectors. That will modify the EAS observables 
($X_{max}$, $RMS$ and $N_{max}$\footnote{$X_{max}$ is the depth of the maximum energy deposited in the atmospheric (starting at
  the top of the atmosphere)  by the EAS, $RMS$ is the fluctuation of the $X_{max}$ (Standard Deviation) and $N_{max}$ is the maximum
  number of particles at the EAS.}), besides a considerable change of the EAS longitudinal profile shape.

In this work, we will use the CORSIKA \cite{Corsika} (Cosmic Ray Simulations for Kaskade) Monte Carlo to simulate the 
evolution of EASs in the atmosphere. We will a modified code of CORSIKA, with charm and bottom production at the 
cosmic ray first interaction \cite{AGascon} (hereafter denoted HQ CORSIKA). In the next section, we will explain how 
this version works. Moreover, the  HQ CORSIKA predictions with those derived using the standard CORSIKA 
(original code - without charm particle production) (hereafter denoted STD CORSIKA). In both versions (HQ CORSIKA
and STD CORSIKA) we will use the QGSJET01 MC  to describe the high energy hadronic interactions and the FLUKA one for the description of the low energy hadronic interactions. We will demonstrate that 
 the charm and bottom production rates are only relevant for the highest energies. The modified MC allows
us to analyse the effects that the production and propagation of heavy hadrons has in the EAS development. 

Our goal is to verify the  impact on the  EAS observables of different selections  of for the minumum value of the $x_F$  variable \footnote{Probability of production of a heavy hadron carrying values larger than a 
  certain fraction of primary energy.}, as well of different models to describe the heavy hadron production  - CGC (Color Glass Condensate) and IQM
(Intrinsic Quark Model).

\section{Heavy particle simulation}

As mentioned earlier, we have included in the CORSIKA MC the production of heavy hadron  (mesons and/or baryons) in the  first interaction of primary cosmic ray with an incident energy $>3\times 10^{17}\;eV$. We consider that 
$\Lambda s, \Xi s, \Sigma s, \Omega s, Ds$ and $Bs$ are produced in this interaction. These particles are produced according
to the probability for the quark charm to be present in different species of hadrons after $p-Air$ collision \cite{Alberto-gap}. 

The CORSIKA MC doesn't explicitly include the charm and bottom heavy interactions.  It is necessary that in the extraction
of the source code  the option CHARM is ``switch on''. Not all packages can handle production and propagation of heavy particles. 
In the CORSIKA, the  DPMJET and QGSJET packages of high energy hadronic interactions consider the interaction of these particles. The $D_s^+$ (main source of tau leptons) particle (PDG code 431) is not included by the QGSJET01. 

We assume that the heavy particles can be produced via two models:

\begin{itemize}
\item Color Glass Condensate - CGC.
\item Intrinsic Quark Model - IQM.
\end{itemize}

The choice of the production model and the kind of heavy particle (charm or bottom) generated in the first 
interaction is made via CORSIKA INPUT - key COLLDR. From the first interaction, the package of hadronic 
interaction PYTHIA \cite{pythia} makes the decay and interaction of the heavy particles. The key PROPAQ 
determines whether the propagation of heavy particles will be handled by PYTHIA, or by standard routine (e.g. QGSJET01). Through key SIGMAQ, the cross sections for the charm and bottom mesons and baryons are determined \cite{AGascon}.

\subsection{Color Glass Condensate}

In this model, a heavy flavor quark-antiquark pair is created through the fluctuaction of a  gluon in the projectile particle.
Charmed and bottom hadrons are formed from the hadronization of these heavy quarks with sea quarks, in a mechanism called
Uncorrelated Fragmentation. More information of the model in \cite{Victor1} and \cite{Victor2}.

When a proton of  energy $E_p$ (in GeV) collides with a nucleus in the atmosphere, the probability to produce a heavy 
hadron carrying a fraction of energy is given by:
\begin{itemize}
\item Above $5\%$ ($x_F>0.05$) \cite{Alberto-gap}:
\end{itemize}
\begin{eqnarray}
P(E_p)=0.00129672 - 0.0000974551*ln(E_p) + 0.000055122*ln(E_p)^2
\end{eqnarray}
\begin{itemize}
\item Above $1\%$  ($x_F>0.01$):
\end{itemize}
\begin{eqnarray}
P(E_p)=-0.0676118 + 0.00544162*ln(E_p) + 0.000166688*ln(E_p)^2
\end{eqnarray}
The energy distribution of charms produced has the general form \cite{Alberto-gap}:
\begin{eqnarray}
\frac{dP}{dx_F}=a*\frac{(1-x_F^{1.2})^{b}}{x_F^{c}}
\end{eqnarray}
where:
\begin{eqnarray}
a=0.0094058+(6.7535\times 10^{-4})*ln(10*E_p),\\
b=8.9416+(-0.02078)*ln(10*E_p),\;\;\;\;\;c=1.3578+(0.01281)*ln(10*E_p) \nonumber
\end{eqnarray}

\subsection{Intrinsic Quark Model}

At leading order in QCD, heavy quarks are produced by the processes $q \overline {q} \rightarrow Q \overline{Q}$
and $g \overline {g} \rightarrow Q \overline{Q}$. When these heavy quarks arise from fluctuation of the initial
state, its wave function can be represented as a superposition of Fock state fluctuations:
\begin{eqnarray}
|h>=c_0|n_v> + c_1|n_vg> + c_2|n_vq \overline {q}> + c_3|n_vQ \overline{Q}> ...
\end{eqnarray}
where $|n_v>$ is the hadron ground state, composed only by its valence quarks.

\begin{figure}[H]
\centering
\hspace{-0.5cm}\includegraphics[width=17.5cm]{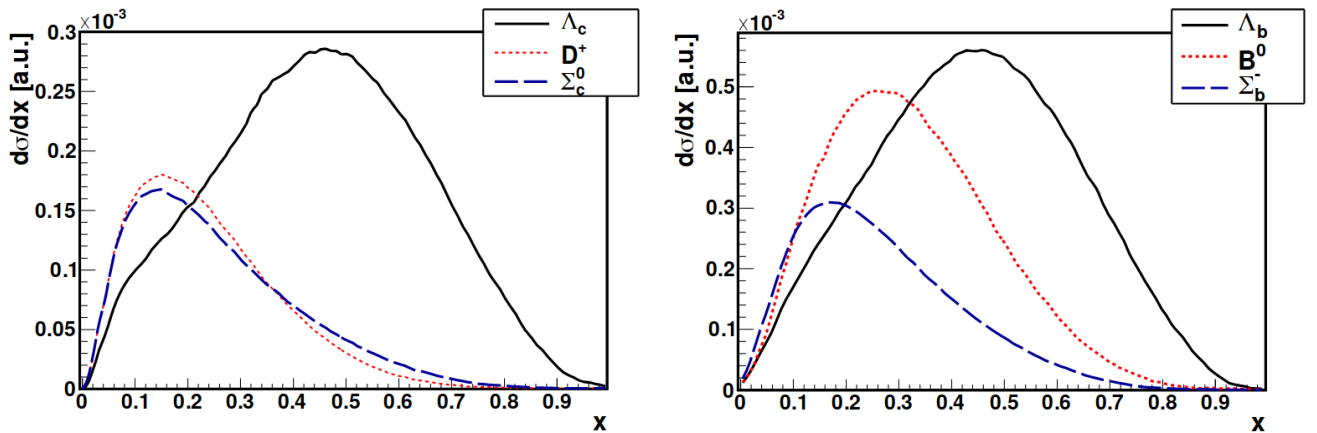}
\vspace{-0.2cm}
\caption{\it Feynman - $x$ distribution  for some charmed (left) and
  bottom (right) hadrons derived using the  Intrinsic Quark production model\cite{AGascon}.}
\label{IQM_diff}
\end{figure}

When the projectile scatters on the target the coherence of the Fock components is broken and the fluctuations can hadronize, either
with sea quarks or with spectator valence quarks. The latter mechanism is called Coalescence. For instance, the production of $\Lambda_c^+$
in $p-N$ collisions comes from fluctuation  of the Fock state of the proton to $|uudc\overline {c}>$ state. To obtain a
$\Lambda_c^-$ in the same collision a fluctuation to $|uudu\overline {u}d \overline {d}c \overline {c}>$ would be required. Thus, since the
probability of a five quarks state is larger than that of a 9 quarks state, $\Lambda_c^+$ production is favored over $\Lambda_c^-$ in proton reactions. 
The co-moving heavy and valence quarks have the same rapidity in these states but the larger mass of the heavy quarks implies they carries most of
the projectile momentum. Heavy hadrons form from these states can have a large longitudinal momentum and carry a large fraction of
the primary energy, which is crucial for their propagation \cite{AGascon}. At the figure \ref{IQM_diff} we can see the differential energy
fraction distribution for some charmed and bottom hadrons.

\vspace{-0.1cm}
The probability to have a state $|uudc\overline {c}>$ in the proton is \cite{IQM1}:
\begin{eqnarray}
P(p \rightarrow uudc\overline {c}) \approx[m_p^2-\sum_{i=1}^5\frac{m_{\perp i}^2}{x_i}]^{-2}
\end{eqnarray}
where the transverse mass is $m_{\perp i}^2$ and we take $i=4,5$ for $c,\overline {c}$.
A detailed description of this model can be found in \cite{IQM1} and \cite{IQM2}.

\section{Longitudinal Profiles}

In this section we will analyse the longitudinal profiles of total energy deposited in the atmosphere, considering $\approx 400$ EAS. We will assume that the primary cosmic ray is $1\times 10^{20}\;eV$, the zenithal angle is $60^\circ$ and primary particle is a  proton. Production of charm
or bottom heavy particles can occur via two models - CGC and IQM. We will use two selections for the minimum of $x_F$ (via CGC) - $x_F>0.01$ and $x_F>0.05$.
The comparisons will be made between CORSIKA HQ and STD CORSIKA. For the longitudinal profile of energy deposited,
the curves will be separated according to the energy fraction ($F_E$) of the  heavy particles produced in the first
interaction - $F_E < 0.1$ and $F_E \ge 0.1$ for the CGC  model and $F_E < 0.5$, $F_E \ge 0.5$ and $F_E \ge 0.8$ for the
IQM production model.

For this analysis, we will restrict the heavy hadron production to $\Lambda_ c^+$, $D^0$, $D^+$, $D_s^+$, $B^+$ and $B^0$ and 
their anti-particles\footnote{Not all high hadronic interactions packages can handle this kind of heavy particles. 
$D_s^+$, $B^+$ $B^0$ and their anti-particles for example is not considered by QGSJET01.}.

\begin{figure}[H]
\centering
\includegraphics[width=10.5cm]{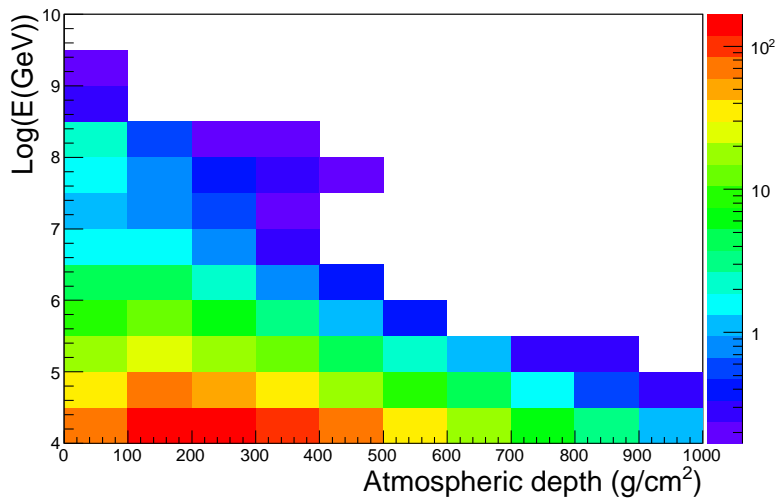}
\vspace{-0.2cm}
\caption{\it Evolution in the production of secondary charms in the EAS for $x_F>0.01$. We assume that the energy of the primary cosmic ray is $3\times 10^{19}\;eV$.}
\label{figura1}
\end{figure}

In Fig.  \ref{figura1} we show the evolution of production of secondary charms in the EAS. The charm  particles
are ``written'' as they are produced during the 
EAS deve\-lo\-pment, from the first interaction of cosmic ray down to the sea level. We assume bins of $100\;g/cm^2$ of atmospheric depths
and half  decade of energy. The  charm particles are dominantly produced with  low  energy (below $10^6\;GeV$) and are  
produced between $100$ and $400\;g/cm^2$. On the other hand, the energetic charms are produced in the first interactions, i.e. when the depth of the atmosphere is  less than 
$200\;g/cm^2$. We have that  $\approx 47$ is the average number of charm produced above $10^6\;GeV$, which will 
decay and produce high energy $\mu$ and $\nu$ that reach the ground. Regarding the   total number of charms produced 
in all bins (energy higher than $10^4\;GeV$), we have an average of $\approx 1000$ charms, being $\approx 400\;D^0$,
$\approx 330\;D^+$, $\approx 47\;D_s$ and $\approx 240\; \Lambda_c$. From this total we have that $\approx 4$ charm are produced with energy higher than $10^8\;GeV$. 

\vspace{-0.4cm}

\begin{figure}[H]
\centering
\includegraphics[width=10cm]{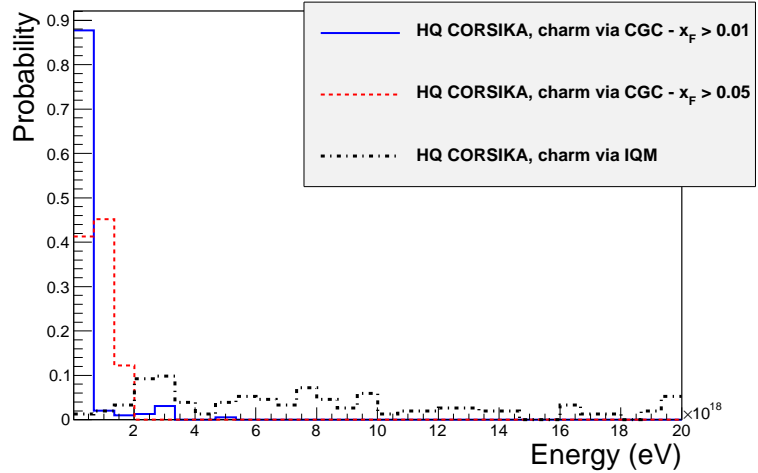}
\vspace{-0.2cm}
\caption{\it Energy distribution of the charm hadrons generated at the first interaction of the HQ CORSIKA. The curves
  are separated according to $x_F$ values ($x_F>0.01$ and $x_F>0.05$) and the model used to describe the heavy quark production.  We assume that the energy of the
  primary cosmic ray is $3\times 10^{19}\;eV$.}
\label{figura2}
\end{figure}

In Fig.  \ref{figura2}, we present the energy distribution for charms produced  in the first interaction obtained using the HQ CORSIKA. 
Comparing the CGC predictions for $x_F>0.01$ and $x_F>0.05$, we have that  the secondary charms have a higher average energy for $x_F>0.05$. However, for $x_F>0.01$ the charms reache higher energies. For example, for $x_F>0.01$ some particles reach $\approx 3\times 10^{18}\;eV$. For 
this energy the charm decay lenght becomes of order of $\approx50\;km$, what could make such particle reach the ground with 
reasonable energy. Comparing now different heavy production models, CGC and IQM, the secondary generated via Intrinsic Quark Model 
reaches a much larger energy than produced via Color Glass Condensate. Via IQM, we have secondary particles been produced at first interaction 
with larger fractions of primary energy. Such particles can carry almost all primary energy.

\vspace{-0.4cm}

Considering now the   energy distribution of the charmed particles that hits the ground. Independently of the selection for the values of  $x_F$ and the production model(CGC and IQM), the number of particles is negligible. Consequently, the most part of charm particles that are produced in the EAS decay or interact before hit the ground.

Muons are a key prediction in EAS simulations. Although the presence of heavy hadrons will not introduce significant differences in the
total number of muons at the ground level, there are other observables that may be more sensitive to these heavy hadrons: Events with
late energy deposition from the decay of a heavy meson or a $\tau$ lepton. The fraction of these events is low \cite{garcia_canal_art_aires}.
Events with leptons of PeV energies, coming from charm decays. 


\vspace{-0.2cm}

\begin{figure}[H]
\centering
\hspace{-0.5cm}\includegraphics[width=9.2cm]{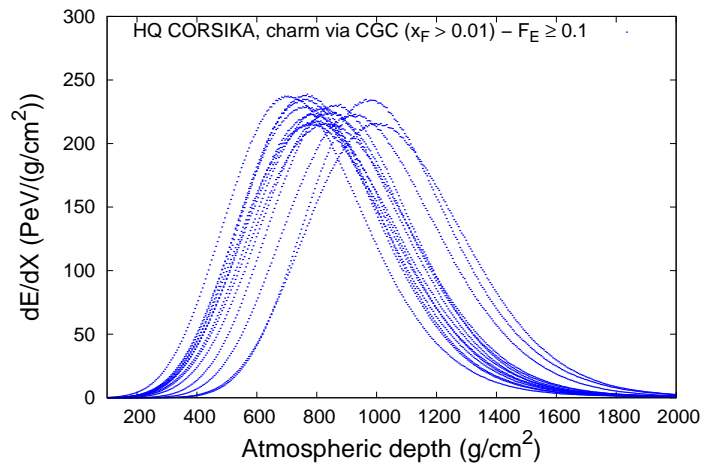}
\vspace{-0.4cm}
\caption{Longitudinal profiles for the total energy deposited in the atmosphere for charm production via CGC ($x_F>0.01$) and $F_E \ge 0.1$.}
\label{figura4}
\end{figure}

\begin{figure}[H]
\centering
\hspace{-0.5cm}\includegraphics[width=9.2cm]{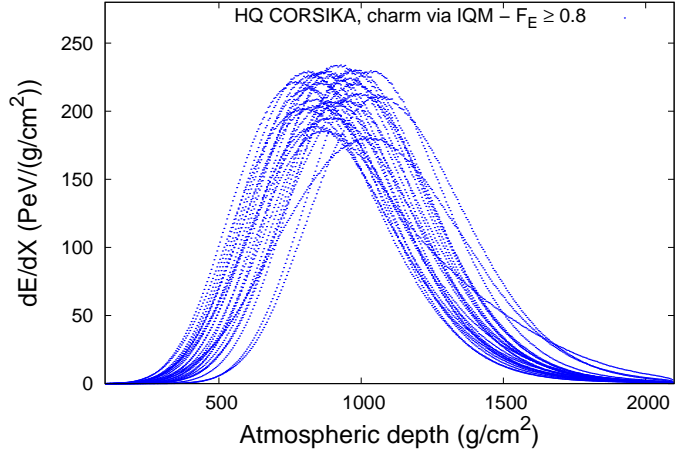}
\vspace{-0.3cm}
\caption{Longitudinal profiles of the total energy deposited in the atmosphere for charm production via IQM, $F_E \ge 0.8$.}
\label{figura5}
\end{figure}

\begin{figure}[H]
\centering
\includegraphics[width=9.2cm]{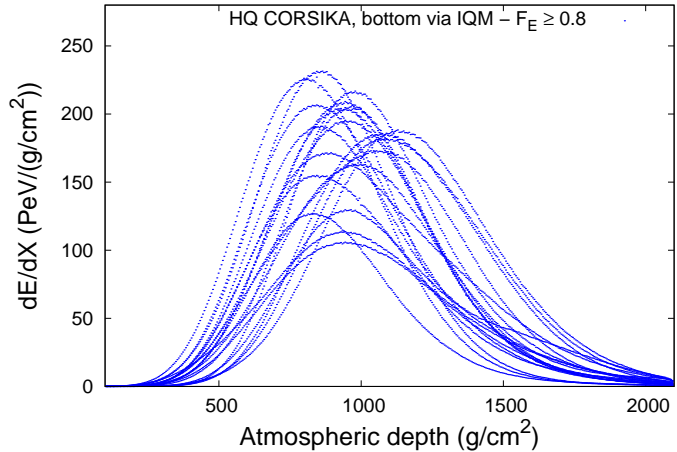}
\vspace{-0.4cm}
\caption{Longitudinal profiles of the total energy deposited in the atmosphere for bottom production via IQM, $F_E \ge 0.8$.}
\label{figura7}
\end{figure}

In the Figs. \ref{figura4}, \ref{figura5}, and \ref{figura7} we present our predictions for the longitudinal profiles.
The longitudinal profiles are separated according to the fraction of the energy of the heavy particles generated in the first interaction,
$F_E < 0.1$ and $F_E \ge 0.1$, which allows to  highlight the profiles. Concerning to $F_E < 0.1$ ($x_F>0.05$), for both charm and bottom production (via CGC),
the longitudinal profiles follow approximately the same behaviour in comparison with the standard profiles (STD CORSIKA).

Concerning to $F_E \ge 0.1$ ($x_F>0.01$, via CGC), presented in the Fig. \ref{figura4}
they represent  less than 6\% of total\footnote{The highest value of the fraction of primary 
energy reached is $\approx 0.2$ (CGC production model), ie, $\approx 6\times 10^{18}\;eV$.}. Looking for the profiles with a fraction $< 0.1$ we 
can't see significant changes in the profile. In fact, they follow the same general shape of profiles with no heavy hadron 
production (STD CORSIKA). If we look now at the profiles with larger fractions, the effect is more
pronounced.  We have a slight difference in the maximum of energy deposited, i.e, we have a shift to deeper layers of the atmosphere. These
profiles have a smaller value for the peak of energy deposited.

\vspace{-0.9cm}


Once again we separated the profiles for $F_E < 0.5$, $F_E \ge 0.5$ and $F_E \ge 0.8$ for the 
IQM heavy production model, considering the charm and bottom production separately. For these cases, we have a large number of profiles with high
energy fraction \footnote{As  mentioned before, heavy hadrons produced via IQM model can have a large longitudinal momentum and carry a large
  fraction of the primary energy.}. For the highlighted profiles with $F_E \ge 0.5$  we predict a  non- negligible change of the longitudinal profiles,
for both particles, charm  and bottom. In particular, for $F_E \ge 0.8$, we have the most important results.
In some profiles the maximum of total energy deposit is largely shifted to larger depths. We also see in some profiles a big change in the profile shape,
i.e, the peak of energy deposited is much smaller\footnote{In some cases the peak of total energy deposited reaches just $100\;PeV/g/cm^2$. The standard
  profile reaches about $250\;PeV/g/cm^2$.} and the profiles are quite elongated (See Figs. \ref{figura5} and \ref{figura7}). This happens because heavy hadrons produced via IQM at first interaction
have high energy fracions, thus carrying a big amount of energy deep into the atmosphere. For  profiles with $F_E \ge 0.8$,  we see significant changes, mainly
for bottom production. In particular, we can see some double core profiles.

\section{Discussions}

Our purpose in this analysis was to study how the presence of a heavy hadrons could modify the  fundamental parameters of the cascade development, 
such as the shape of longitudinal profile, the number of particles reaching the ground, the position and deviation of the 
shower maximum. Heavy hadrons propagating with an energy above its critical value will travel long paths. In particular, 
showers with zenithal angle of $60$ degrees have an atmosphere slant depth of $\approx 2100\;g/cm^2$. After several elastic 
interactions, we expect heavy particles to deposit its remaining energy deep in the atmosphere and some could reach 
the ground carrying substancial energy fraction. If the heavy hadron carries a significant fraction 
of the primary's energy, we can expect a large impact on the EAS observables. As the heavy particle energy fraction
increases, more accentuated these effects will be.


\begin{figure}[H]
\centering
\includegraphics[width=10.5cm]{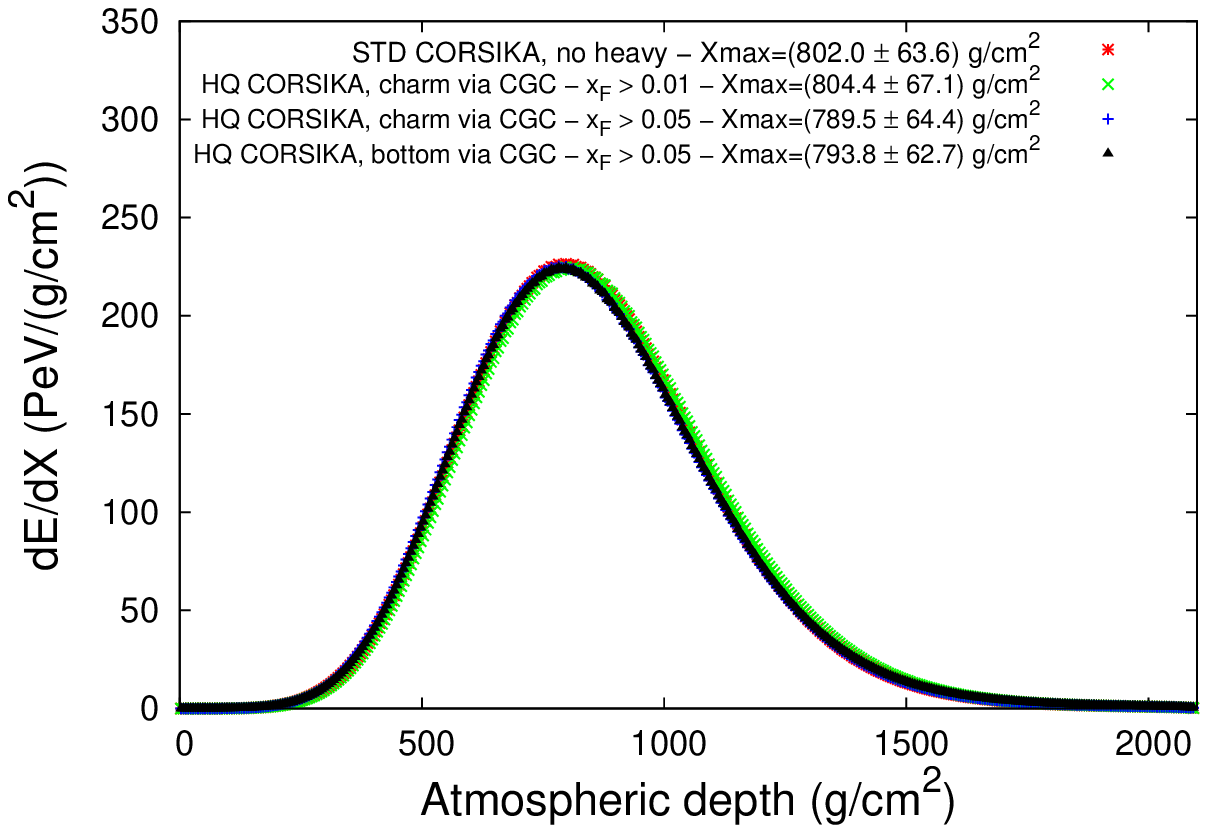}
\includegraphics[width=10.5cm]{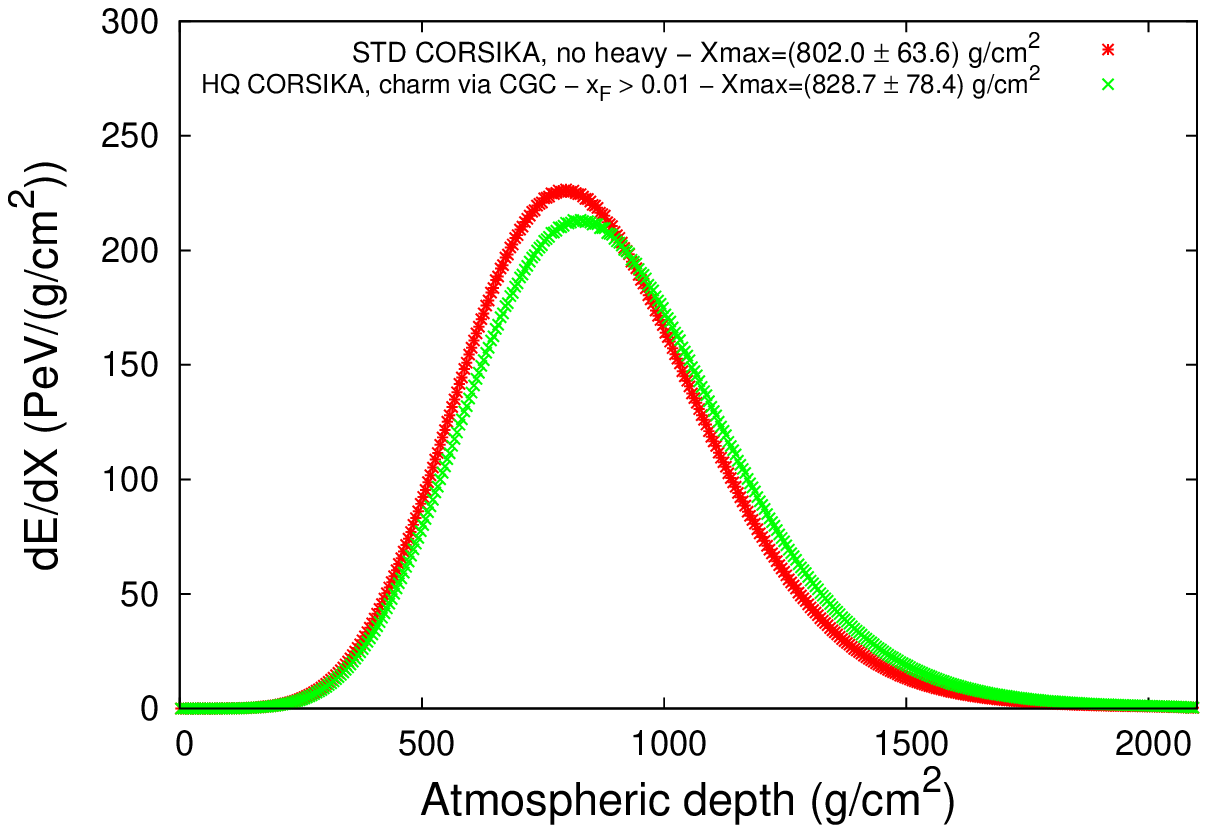}
\vspace{-0.2cm}
\caption{Longitudinal profiles of the energy deposited in the atmosphere considering the  HQ CORSIKA (CGC, to $x_F>0.01$ 
and $x_F>0.05$) and the STD CORSIKA. Upper panel: Predictions for $F_E< 0.1$. Lower panel: $F_E\ge 0.1$.}
\label{figura8}
\end{figure}

In what follows we will present our predictions for the average longitudinal profiles. In Fig. \ref{figura8}, we can't observe significant differences between the  average profiles. Small changes of the 
observable occur only for  $x_F>0.01$ ($F_E \ge 0.1$). In this case the $X_{max}$ is more deeper and the $RMS$ larger in comparison with EAS 
simulated by the STD CORSIKA. We have a relative difference of  9\% of to $X_{max}$ and 28\% to $RMS$. Comparing now the $x_F$ assumed, 
we have small differences in the shape of longitudinal profiles. For $x_F>0.01$, the $X_{max}$ and $RMS$ is slightly larger.

\begin{figure}[H]
\centering
\includegraphics[width=11.5cm]{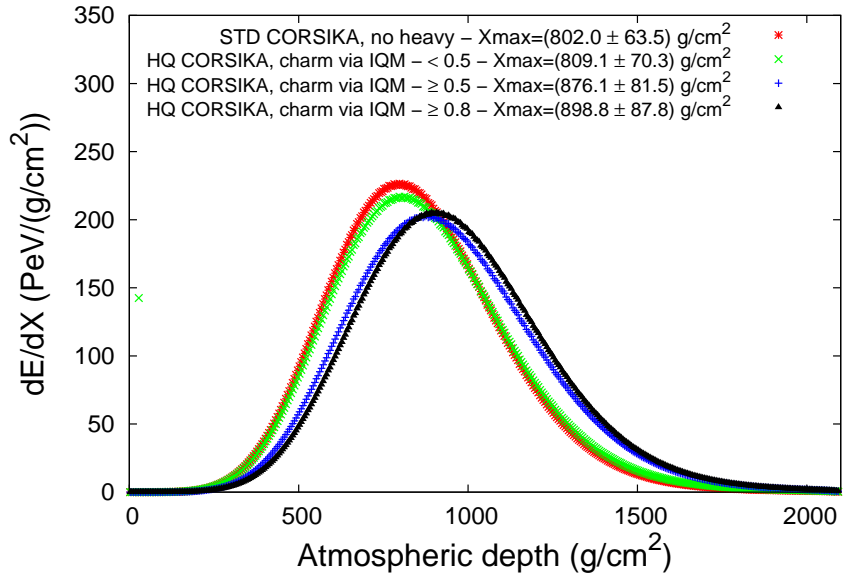}
\vspace{-0.4cm}
\caption{Longitudinal profiles of the energy deposited in the atmosphere considering  HQ CORSIKA (IQM - charm production) and the 
STD CORSIKA. Predictions for  $F_E< 0.5$, $F_E\ge 0.5$ and $F_E\ge 0.8$.}
\label{figura9}
\end{figure}

\begin{figure}[H]
\centering
\includegraphics[width=12cm]{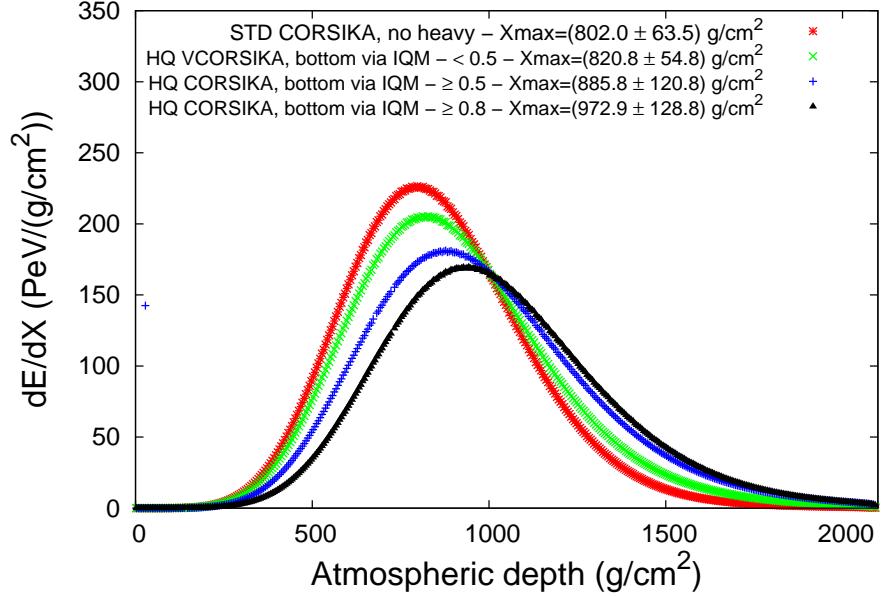}
\vspace{-0.4cm}
\caption{Longitudinal profiles of the energy deposited in the atmosphere considering the HQ CORSIKA (IQM - bottom production) and the
STD CORSIKA. Predictions for $F_E< 0.5$, $F_E\ge 0.5$ and $F_E\ge 0.8$.}
\label{figura10}
\end{figure}

Regarding the average profiles obtained using  IQM production model,  presented in Figs.  \ref{figura9} and \ref{figura10}, we can see significant changes in the 
longitudinal profiles. We have  a deeper $X_{max}$ and a higher $RMS$ when the  energy fraction of the heavy particles is increased. For charm production
with $F_E \ge 0.8$ we predict the values of $898.8\;g/cm^2$ and $87.8\;g/cm^2$, respectively, for $X_{max}$ and $RMS$. For bottom production with $F_E \ge 0.8$
we obtain $972.9\;g/cm^2$ and $128.8\;g/cm^2$.
We can also observe a significant change in the shape of the profile. Depositing less energy according to energy fraction. Bottom and charm particle 
interaction are more elastic than other particles, therefore charms and bottoms produced with high primary fraction will deposit energy more slowly in 
the atmosphere and can carry large energies deeper in the atmosphere. Such effect is larger in bottom particles. At higher fractions of
energy  ($F_E \ge 0.8$), the impact on the average longitudinal profile is larger. In the case of charm production we have a $X_{max}$ shifted to deeper layers of the atmosphere in relation to standard CORSIKA, with the relative difference 
being about 12\%. For the $RMS$, we have a relative difference of  40\%. Regarding bottom production we have a more radical 
effect, being 22\% for $X_{max}$ (shift to deeper layers) and 100\% to $RMS$. For the RMS, we have larger values,
which means that the fluctuation of depth of the maximum of EAS ($X_{max}$)  is larger. This happens because the heavy particle interaction is more elastic.
All these EAS effects ($X_{max}$ shift, larger $RMS$ and more elongated shape of the EAS profiles) can change significantly the EASs longitudinal
profile seen in fluorescence telescopes and/or the temporal distribution observed in the surface detectors. In fact, we could see some double core
profiles. The global effect of all these changes in the longitudinal profile is the lower energy reconstructed of the EAS and higher uncertainties. For
the energy of the primary cosmic rays considered ($\approx 10^{20}\;eV$), the values of $X_{max}$ and $RMS$ found in experiments such as the Pierre Auger
Observatory for example are respectively\footnote{Here it is taken into account that we have a mixture of several mass compositions (H, He, C, Fe, etc.)}
$\approx 760\;g/cm^2$ and $\approx 26\;g/cm^2$ \cite{PRD90}. The $X_{max}$ and $RMS$ are directly linked to the mass composition of the primary cosmic ray.
The appearance of charm and bottom in the EAS makes it more difficult to make such a connection, because of the deeper $X_{max}$ and larger $RMS$. 

\begin{figure}[H]
\centering
\includegraphics[width=11.5cm]{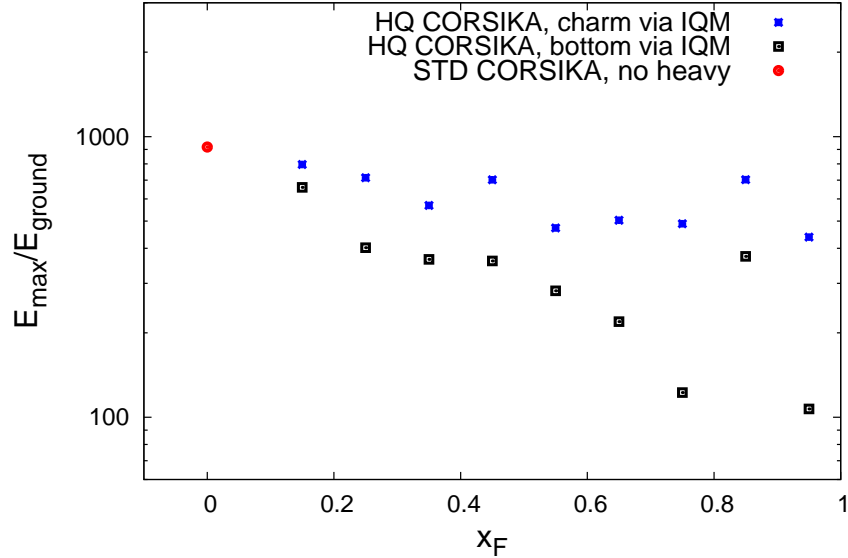}
\vspace{-0.4cm}
\caption{Ratio between the number of particles at EAS maximum and the number of particles at ground level according to fraction of 
energy carried by the heavy hadron. The red point it related to ratio predicted by the STD CORSIKA.}
\label{figura11}
\end{figure}

In Figs. \ref{figura9} and \ref{figura10} we can also see a more elongated shape of the EAS, ie, 
a slower energy deposit. In these figures we analyze when the energy deposition in the shower is shifted to large depths. 
The amplitude and position of EAS $X_{max}$ is affected. The number of particles at maximum  decrease, while 
the number of particles that reach the ground increase. In Fig. \ref{figura11}, we show the ratio between the 
number of particles in the maximum of the shower and the number of particles at the ground level ($E_{max}/E_{ground}$) according to the fraction 
of energy carried by the heavy hadron. Such ratio is sensitive to the change in the profile's shape. The EAS energy deposited 
amplitude decrease as the $X_{max}$ is shifted to higher depths. The effect is higher to bottom particle production because of 
its higher elasticity.

\begin{figure}[H]
\hspace{-0.5cm}\includegraphics[width=8.7cm]{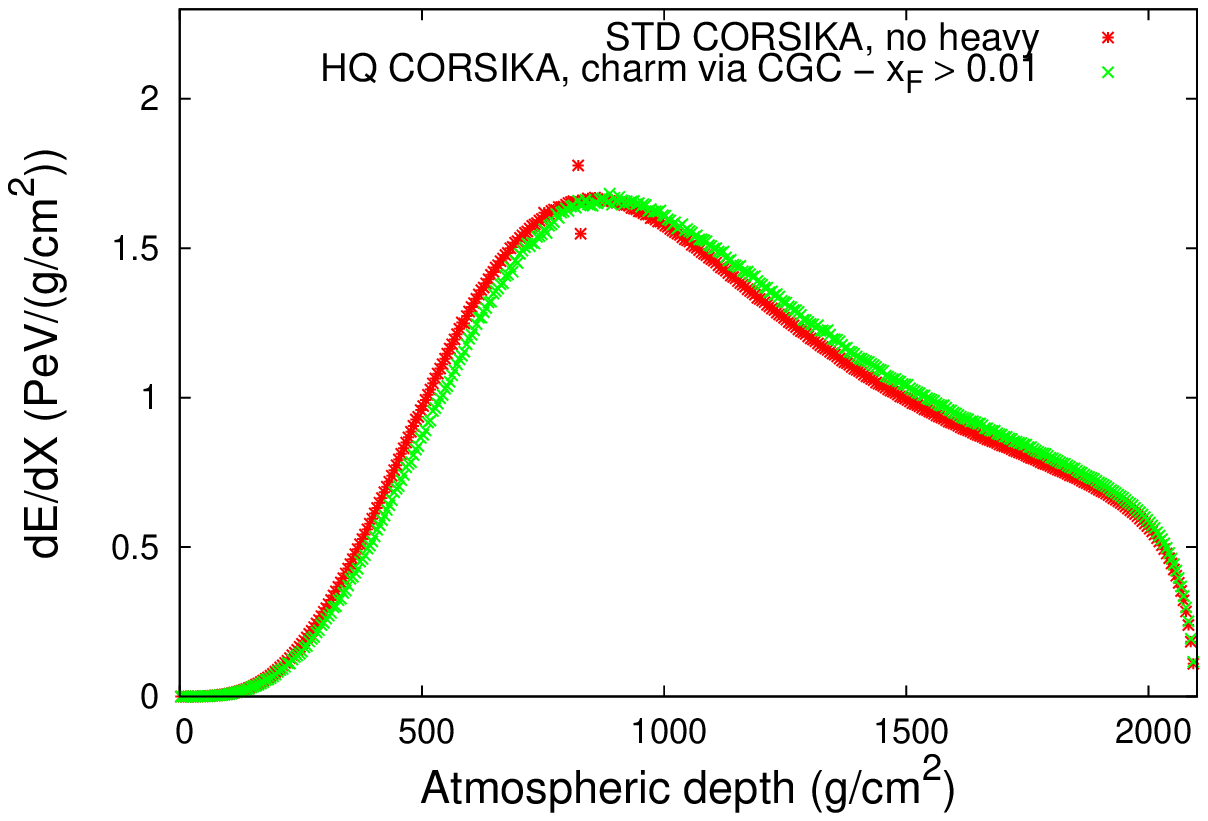}
\hspace{-0.4cm}\includegraphics[width=8.7cm]{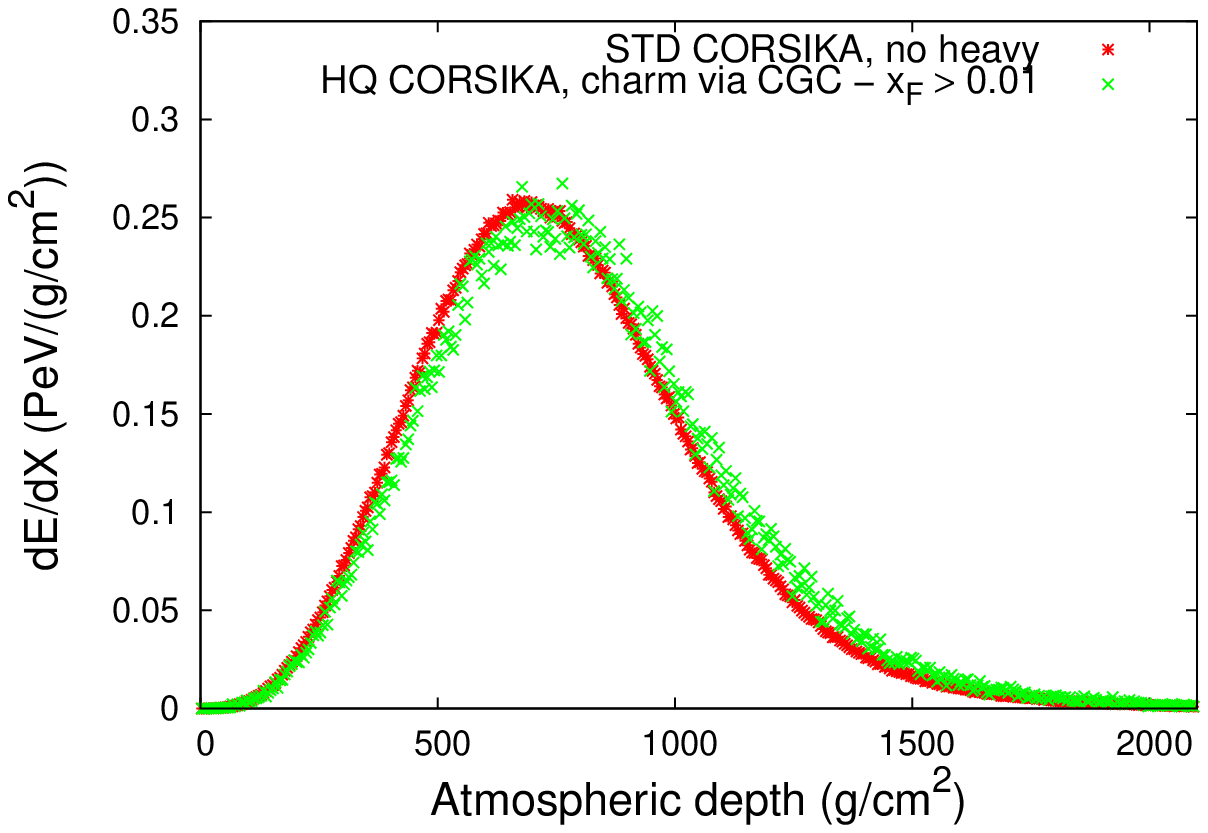}
\hspace*{-0.5cm}\includegraphics[width=8.7cm]{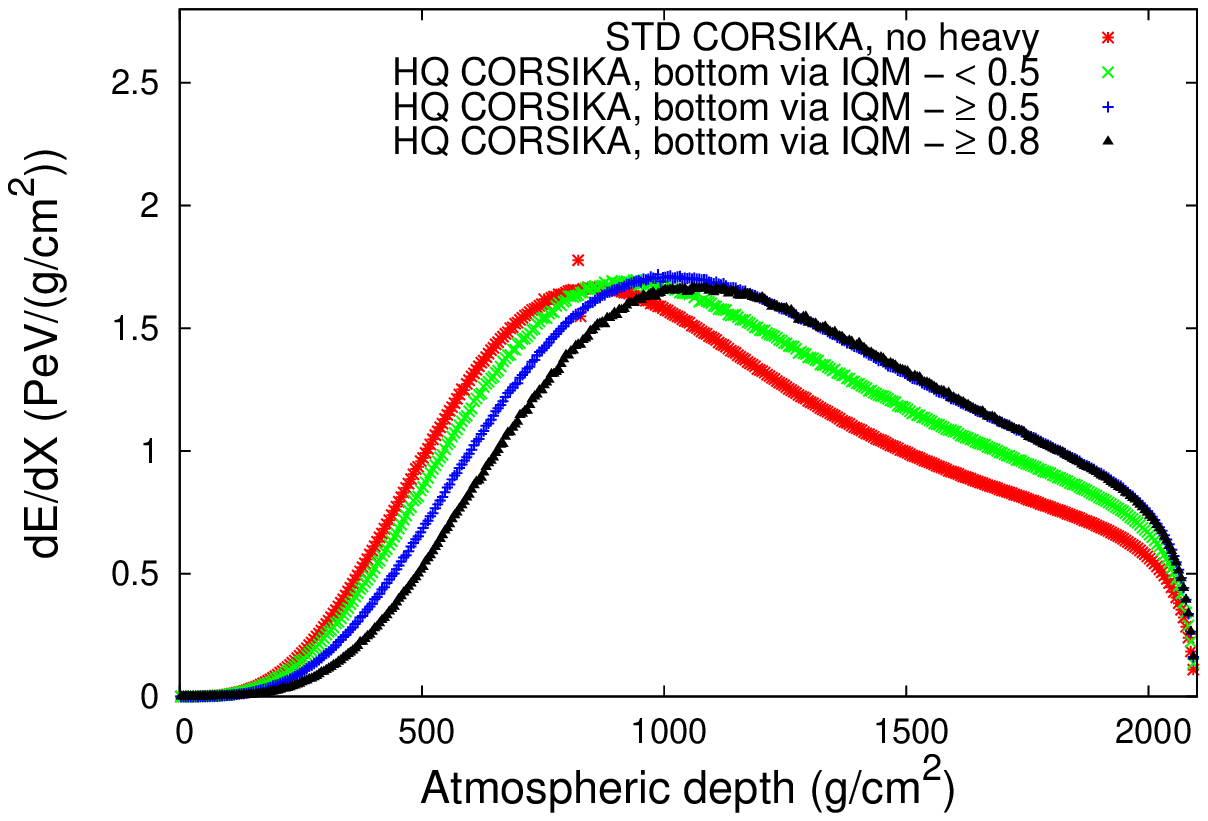}
\hspace{-0.0cm}\includegraphics[width=8.7cm]{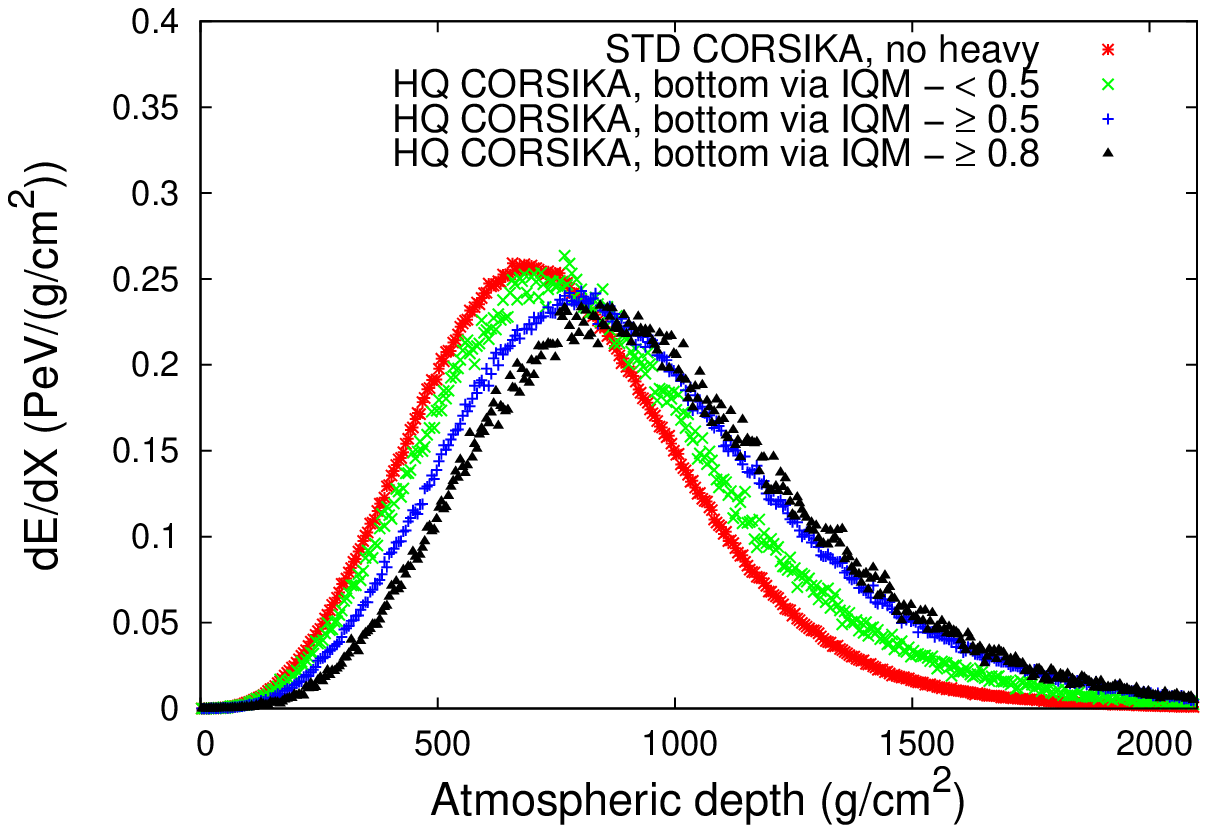}
\vspace{-0.3cm}
\caption{Longitudinal profiles of the  energy deposited in the atmosphere by muons (left) and neutrinos (right) considering  the HQ CORSIKA (CGC, to $x_F>0.01$, and IQM) and the STD CORSIKA. Above, 
$F_E\ge 0.1$. Below, we have $F_E< 0.5$, $F_E\ge 0.5$ and $F_E\ge 0.8$.}
\label{figura12}
\end{figure}

In Fig. \ref{figura12} we present the average longitudinal profiles for  the energy deposited  by muons and neutrinos. The shape of curves is shifted to IQM bottom production, both for muons an neutrinos. Again, this effect is 
more significant when we consider higher energy primary fractions.

\section{Conclusions}

Regarding average longitudinal profile via CGC, we observe small changes in the $X_{max}$, $RMS$ and $N_{max}$ 
observables. We can see small effects only for  individual profiles. The energy of the heavy secondaries produced via CGC reaches up to $\approx 3\times 10^{18}\;eV$, what could make such 
particles reach the ground with reasonable energy. However,  the fraction of energy carried by the particles 
is very small to cause considerable effects in the EAS development. 

Regarding average longitudinal profiles via IQM, the heavy particles reach  higher energy fraction, causing 
larger changes in the  observables. We observed a considerable change in $X_{max}$, $RMS$ and $N_{max}$. We highlight 
some longitudinal profiles with high larger modification in its shape. In fact we can observe some double core profiles. This behavior 
certainly will cause important effects in EAS detection. Charm and bottom particles are very rare in EAS, but we 
shown that its effects are radical in the EAS development. All these EAS effects ($X_{max}$ shifted, larger $RMS$ and more elongated
shape of the EAS profiles) can change significantly the EASs longitudinal profiles seen in fluorescence telescopes and/or the temporal
distribution observed in the surface detectors. The global effect of all these changes in longitudinal profile is a smaller value of the energy
reconstructed of the EAS and higher uncertainties. As for the energy of the primary cosmic rays in question ($\approx 10^{20}\;eV$), the values
of $X_{max}$ and $RMS$ found in experiments such as the Pierre Auger are smaller. The $X_{max}$ and $RMS$ are directly linked to the mass
composition of the primary cosmic ray. The appearance of charms and bottoms in the EAS makes it more difficult to make such a connection,
because of the deeper $X_{max}$ and larger $RMS$. The discussion of whether the detection of EAS with heavy particles is feasible in
Fluorescence telescopes or Surface Detectors needs to be analyzed in more detail and will be done in a future publication.    

The inclusion of heavy hadrons in CORSIKA opens the possibility to test new possibilities of theories to heavy 
hadron production and propagation.

\vspace{0.7cm}

\begin{center}
{\Large \bf Acknowledgments}
\end{center}

\vspace{0.7cm}

We would like to thank Alberto Gascón for useful discussions. This work was  partially financed by the Brazilian funding
agencies CNPq, CAPES,  FAPERGS and INCT-FNA (process number 
464898/2014-5).


\begin{thebibliography}{10}

\bibitem{garcia_canal_art_aires} C. A. García, et al, {\it Production and propagation of heavy hadrons in air-shower simulators}, 
Astroparticle Physics, 46, 29-33, 2013.
\bibitem{Auger} http://www.auger.org/
\bibitem{Corsika} Heck, D., Knapp, J., Capdevielle, J. N., Schatz, G., and Thouw, T., Report  FZKA 6019 (1998), 
Forschungszentrum Karls\-ruhe, Germany.
\bibitem{AGascon} A. Bueno and A. Gascón, {\it Corsika implementation of heavy quark production and propagation 
in Extensive Air Showers}, Computer Physics Communications, 185, 638-650, 2014.
\bibitem{pythia} http://home.thep.lu.se/~torbjorn/Pythia.html
\bibitem{Alberto-gap}  A. Gascón and A. Bueno, {\it Charm production and identification in EAS}, Gap Note 
(Internal notes of Auger Collaboration), 2011-019, 2011.
\bibitem{Victor1} V. P. Gonçalves and M. V. T. Machado, {\it Saturation physics in ultra high energy cosmic rays: heavy
  quark production}, Journal of High Energy Physics, 04:028, 2007. 
\bibitem{Victor2} E. R. Cazaroto, V. P. Gonçalves and F. S. Navarra, {\it Heavy quark production at LHC in the color dipole formalism},
  Nuclear Physics A, 872(1):196-209, 2011.
\bibitem{IQM1} N. Sakai, P. Hoyer, C. Peterson and S. J. Brodsky, {\it The intrinsic charm of the proton}, Phys. Lett. B, 93:451-455, 1980. 
\bibitem{IQM2} R. Vogt, {\it Charm Production in Hadronic Collision}, Nuclear Physics A, 553:791-798, 1993.
\bibitem{PRD90} Pierre Auger Collaboration, {\it Depth of Maximum of Air-Shower Profiles at the Pierre Auger Observatory: Measurements at Energies above $10^ {17.8}\;eV$},
  Phys. Rev. D90, 12, 122005, 2014.
\end{thebibliography}
\end{document}